\documentclass[12pt]{article}
\usepackage{graphicx}% Include figure files
\newcommand{\sss}{\scriptscriptstyle}

\newcommand {\be}{\begin{equation}} % start equation
\newcommand{\ee}{\end{equation}}    % end equation

\def\dds1{\frac{\partial}{\partial s_1}}

\def\vte{v_{{\sss T}e}}

\def\d{d\kern-0.8 ex\vrule height 1.3 ex depth-1.24 ex width 0.7 ex
\kern 0.15 ex}
\def\D{D\kern-1.7 ex\vrule height .87 ex depth-0.8 ex width 0.7 ex
\kern 0.95 ex}

\evensidemargin
 0.3cm
\begin{document}
\baselineskip 20 pt

\begin{center}

\Large{\bf Perpendicular electron collisions in drift and acoustic
wave instabilities}

\end{center}

\vspace{0.7cm}

\begin{center}

 J. Vranjes and S. Poedts

{\em K. U. Leuven, Center for Plasma Astrophysics, Celestijnenlaan
200B, 3001 Leuven,
 Belgium, and Leuven Mathematical Modeling and Computational Science Center
 (LMCC)}

\end{center}

\vspace{2cm}

{\bf Abstract:} Perpendicular electron dynamics and the associated
collisions are discussed in relation to the collisional drift wave
instability. In addition, the limit of small parallel wave numbers
of this instability is studied and it is shown to yield a reduced
wave frequency. It is also shown that in this case the growth rate
in fact {\em decreases} for smaller parallel wave numbers, instead
of growing proportional to $1/k_z^2$. As a result, the growth rate
appears to be angle dependent and to reach a maximum for  some
specific direction of propagation.  The explanation for this strange
behavior is given. A similar analysis is performed for acoustic
perturbations in plasmas with unmagnetized ions and magnetized
electrons, in the presence of a density gradient.

\vspace{2cm}

\noindent PACS Numbers: 52.35.Kt, 52.30.Ex, 52.20.Fs, 52.35.Fp

\vspace{2cm}

\pagebreak

%\paragraph{Introduction and Model.}

\vspace{0.7cm}

\noindent{\bf I. \,\,\, INTRODUCTION  AND MODEL}

\vspace{0.7cm}

Typically, collisions in plasmas are responsible for the damping of
plasma modes. However, this does not hold for the drift wave which
grows in the presence of electron collisions. The growth appears
because, in the presence of collisions, the perturbed wave potential
lags behind the perturbed density. This, together with the free
energy stored in the density gradient, makes the mode growing.  In
the case when the ions are not magnetized (e.g., perturbations with
frequencies above the ion gyrofrequency $\Omega_i$, or/and when the
ion collision frequency is above $\Omega_i$), the ion motion will be
of the  sound-type, while  the motion of electrons can still be
described within the drift wave limits. Hence, acoustic-type
perturbations in such a case may also become unstable due to the
same reasons,$^{1,2}$ with the instability taking place provided
that the acoustic wave frequency is {\em below} the electron
diamagnetic frequency $\omega_{*e}= v_{*e} k_y$, $v_{*e}=-
(\vte^2/\Omega_e) \vec e_z \times \nabla n_0/n_0$, $\vte^2= \kappa
T_e/m_e$. Physically, the instability  is due to the fact that
collisions prevent the electrons from  moving freely and from
shielding the electrostatic perturbations that involve  the ion
component.

In most cases,  these electron collision effects are included from
the electron parallel momentum.$^{3-6}$  This is because of the
small inertia, which results in  a dominant electron motion in the
parallel direction, as compared to the perpendicular
gyromotion-type dynamics. Yet, a clear distinction about when such a
model is justified is rarely seen  in the literature. In our recent
work$^{7}$ it is shown that the model can be used  provided  that
\be k_z^2 \Omega_e^2/(k_y^2 \nu_e^2) \gg 1. \label{e1}
 \ee
Here, $k_z$ and $k_y$ are the wave number components  in the
direction parallel and perpendicular to the direction of the
magnetic field vector $\vec B_0=B_0 \vec e_z$, respectively,  and
$\Omega_e$ denotes the electron gyrofrequency. The electron
collision frequency $\nu_e$ may include collisions with both ions
and neutrals. Numerous parameters contribute to the condition
(\ref{e1}), and consequently it may not always be satisfied, so that
in some cases it may become necessary to include  collisions  in the
perpendicular electron dynamics as well. An obvious consequence of
these  collisions  is that the usual  drift motion of electrons
(perpendicular to the force vector) will be modified, and an
additional component of  motion in the direction of force vector
will appear as well.

In a weakly ionized plasma, the electrons may predominantly collide
with neutrals. Usually, these collisions are described assuming a
static and heavy neutral background so that the dynamics of the
neutrals is ignored. However, the interaction between the two fluids
happens due to friction, and the momentum conservation imposes that
the neutral dynamics must be taken into account in spite of a big
difference of mass (per unit volume) of the two fluids. In a recent
study$^{6}$  the differences between the two models are discussed in
detail.

In the present work, we consistently take the dynamics of the target
particles (neutrals in the present case) into account, and we
discuss the domain in which the condition (\ref{e1}) is not
satisfied.

In the case when the temperature of the heavy plasma constituents
(ions and neutrals) is much below the electron temperature, we omit
their thermal effects. The electron momentum equation is
\be
m_en_e \left[\frac{\partial \vec v_e}{\partial t} + (\vec
v_e\cdot\nabla)\vec v_e\right] = e n_e\nabla \phi  - e n_e \vec
v_e\times \vec B
 - \kappa T_e \nabla n_e
- m_e n_e \nu_{en}(\vec v_e - \vec v_n), \label{e2} \ee and for the
neutrals we use
\be
 \left[\frac{\partial}{\partial t} + (\vec
v_n\cdot\nabla)\right] \vec v_n= - \nu_{ne}(\vec v_n-\vec v_e).
\label{e3} \ee
The momentum conservation implies that $\nu_{ne}= m_e n_e
\nu_{en}/(m_n n_n)$.

\vspace{0.7cm}

\noindent{\bf II. \,\,\, ON THE EQUILIBRIUM}

\vspace{0.7cm}

%\paragraph{On equilibrium.}
It is seen that, assuming a stationary equilibrium and also $(\vec
v_{n0}\cdot\nabla)\vec v_{n0}=0$,  $(\vec v_{e0}\cdot\nabla)\vec
v_{e0}=0$, from (\ref{e3}) we have $\vec v_{n0}=\vec v_{e0}$. At the
same time, Eq.~(\ref{e2}) yields $\vec v_{e0}= \vec v_{*e}$. Hence,
due to friction both fluids move together. In most cases under
laboratory conditions such a state is not expected to take place
because of the long time that is needed to reach it. In order to get
a better feeling about the necessary time scales, we may set a
constant velocity for the electrons $\vec v_{e0}$ (assuming that the
parameters determining the velocity are controlled externally).
Using Eq.~(\ref{e3}) we then obtain the neutral velocity: $
v_{n0}/v_{e0}=1-\exp(-\nu_{ne} t)$. Now, setting for example
$n_{e0}=10^{16}$ m$^{-3}$,  and assuming a hydrogen plasma in
hydrogen gas with $n_{n0}=10^{19}$ m$^{-3}$, $T_e=1$ eV, we have
$\nu_{en}=1.09 \cdot10^6$ Hz, $\nu_{ne}=0.6$ Hz, and we find that $
v_{n0}/v_{e0} \simeq 1$ within about 8~s. Here $\nu_{en}=
\sigma_{en} n_{n0} \vte$, and$^{8}$ $\sigma_{en}(T_e) \simeq 2.6
\cdot 10^{-19}$ m$^2$.

Similarly, setting $n_{e0}=10^{17}$ m$^{-3}$,  $n_{n0}=10^{19}$
m$^{-3}$, $T_e=5$ eV,  we have $\nu_{en}=1.07 \cdot 10^6$ Hz,
$\nu_{ne}=5.8$ Hz, and $ v_{n0}/v_{e0} \simeq 1$ within about 0.8~s.
Here$^{8}$  $\sigma_{en}(T_e) \simeq 1.14 \cdot 10^{-19}$ m$^2$.

Numbers used here are for demonstration only. Yet, similar
parameters may be seen in various studies dealing with collisional
drift-waves, like in the experimental work from Ref.~9 (with $T_e=2$
eV, $T_i\leq 0.1$ eV,  $n_{e0}=10^{16}$ m$^{-3}$ and
$\nu_{en}=2\cdot 10^6$ Hz). A similar density ($n_{e0}=7\cdot
10^{16}$ m$^{-3}$) but a 5 times lower electron temperature is used
in the experiment in Ref.~10, while in the experiment in Ref.~11 the
density and the temperature are  also of the same order as in our
examples, i.e., $n_{e0}=5 \cdot 10^{15}$ m$^{-3}$, $T_e\simeq 1$ eV.
In another experiment$^{12}$ the parameters are $n_{e0}=2 \cdot
10^{15}$ m$^{-3}$, $T_e=2$ eV.

Although parameters may have very  different values in various
experiments, the number used in the examples above are frequently
seen in the laboratory plasmas. Although the ionization degree in
the two examples given above is not so small, the electron-neutral
collisions are larger than electron-ion collisions. For the given
parameters, the two cases yield (in kHz) $\nu_{ei}\simeq 304, \, 305
< \nu_{en}$, respectively. Here $\nu_{ei}$ remains almost the same
because both the density and the temperature are increased. We
used$^{13,14}$ the expression for the collision frequency between
any two charged species $b$ and $a$
 \be
 \nu_{ba}= 4\left(\frac{2\pi}{m_b}\right)^{1/2}
  \left(\frac{q_aq_b}{4 \pi \varepsilon_0}\right)^2 \frac{n_a L_{ba}}{3(T_b +  T_a m_b/m_a)^{3/2}}.
   \label{cf}
  \ee
Here, $L_{ba}=\log(\lambda_d/b_0)$ is the Coulomb logarithm,
$\lambda_d=\lambda_{db} \lambda_{da}/(\lambda_{db}^2+
\lambda_{da}^2)^{1/2}$ is the plasma Debye radius, and $b_0=[|q_b
q_a|/[12 \pi \varepsilon_0 \kappa (T_b+ T_a)]$ is the impact
parameter for Coulomb collisions.

Hence, it is correct to say that in most cases we have a `time
evolving  background plasma'. Yet, this time dependence is
negligible as long as the period of the perturbations is much
shorter than the equilibrium evolution time (that is of the order of
$1/\nu_{ne}$). In fact, this is not a rough approximation bearing in
mind the other plasma parameters and the corresponding evolution.
For example, the assumed equilibrium density gradient is also time
dependent because of the electron collisions, and those take place
on much shorter time scales. In other words, fast electron
collisions will tend to destroy the density gradient, and this may
happen  on  short time scales, unless the density gradient is
somehow kept externally (examples of that kind can be found in
Refs.~15,~16). This can be seen by taking the second set of the
above given parameters and assuming $B_0=0.01$ T, and the  plasma
density gradient length as $L_n  \equiv
[(dn_{e0}/dx)/n_{e0}]^{-1}=0.1$ m. The perpendicular electron
diffusion velocity is then given by $D_e/L_n=3$ m/s, where
$D_e=\nu_{en} \vte^2/\Omega_e^2$. Hence, electron diffusion
perpendicular to the magnetic field vector will tend to destroy the
given density gradient in about 0.03~s and all this is entirely due
to collisions. Regarding the neutrals, we stress that there may also
be a continuous diffusion of neutrals from nearby regions, and those
new particles  have no time to be accelerated by friction within the
assumed short wave period. As a result, in most cases one may
perform a wave analysis neglecting the equilibrium movement of the
neutrals.

\vspace{0.7cm}

\noindent{\bf III. \,\,\, DERIVATIONS AND RESULTS }

%\vspace{0.7cm}
%\vspace{0.4cm}

\noindent{\bf A. Drift wave}

\vspace{0.4cm}

%\paragraph{Derivations.}
From Eq.~(\ref{e2}), in the limit when the wave phase velocity and
the perturbed electron velocity are much below the electron thermal
velocity (equivalent to the usual massless electron approximation in
the literature), we obtain the electron  perpendicular velocity \be
v_{e\bot}= \frac{1}{1+ \nu_{en}^2
\alpha^2/\Omega_e^2}\left[\frac{1}{B_0}\vec e_z\times \nabla_\bot
\phi +\frac{\nu_{en} \alpha}{\Omega_e} \frac{\nabla_\bot \phi}{B_0}
- \frac{v_{\sss{T} e}^2\nu_{en} \alpha}{\Omega_e^2}
\frac{\nabla_\bot n_e}{n_e}  -  \frac{v_{\sss{T} e}^2}{\Omega_e}
\vec e_z\times \frac{\nabla_\bot n_e}{n_e}\right].  \label{e5} \ee
Here, $\alpha=\omega/(\omega + i \nu_{ne})$. In view of the previous
discussion on equilibrium, the perturbed neutral velocity used here
is $\vec v_n=i \nu_{ne} \vec v_e/(\omega + i \nu_{ne})$. The
electron parallel momentum yields the parallel  velocity \be
v_{ez1}=\frac{i k_z \vte^2}{\nu_{en}} \frac{\omega^2 +
\nu_{ne}^2}{\omega^2 - i \nu_{ne} \omega} \left(\frac{e
\phi_1}{\kappa T_e} - \frac{n_{e1}}{n_{e0}} \right). \label{e6} \ee
Using Eqs.~(\ref{e5},~\ref{e6}), and neglecting $|\nu_{en}^2
\alpha^2/\Omega_e^2|$ in comparison to 1 in the denominator in
Eq.~(\ref{e5}) (the limit  of magnetized electrons), from  the
electron continuity equation
\[
\frac{\partial n_{e1}}{\partial t}+ \nabla_\bot (n_e \vec v_{\bot
e}) + \nabla_z (n_{e0} \vec v_{ez1})=0,
\]
we obtain \be
\frac{n_{e1}}{n_{e0}}= \frac{\omega_{*e} + i D_p + i D_z (\omega^2 +
\nu_{ne}^2)/(\omega^2- i \nu_{ne} \omega)}{ \omega  + i D_p + i D_z
(\omega^2 + \nu_{ne}^2)/(\omega^2- i \nu_{ne} \omega)} \frac{e
\phi_1}{\kappa T_e}. \label{e7} \ee Here
\[
  D_p= \nu_{en}\alpha k_y^2 \rho_e^2, \quad D_z= k_z^2 \vte^2/\nu_{en}, \quad \rho_e=\vte/\Omega_e.
 \]
The term $D_p$ describes the effects of electron collisions in the
perpendicular direction. Note that neglecting the neutral dynamics
here is equivalent to setting $\nu_{ne}=0$, yielding $\alpha=1$, and
this corresponds to the model in Refs.~1,~2. In the case $D_p=0$,
Eq.~(\ref{e7}) becomes the same as the corresponding equation from
Ref.~6. It is seen also that the assumption  $|D_z/D_p|\gg 1$ yields
the condition (\ref{e1}). In the present work we are interested in
the domain when this limit is either reversed or the ratio is of
order of 1.

For cold collisionless ions we use the same model as in Ref.~6,
i.e., the ions are only subject to the electromagnetic  force, so
that the momentum equation for magnetized ions is \be
m_in_i \left[\frac{\partial \vec v_i}{\partial t} + (\vec
v_i\cdot\nabla)\vec v_i\right] = -e n_i\nabla \phi  + e n_i \vec
v_i\times \vec B. \label{i1} \ee Using this, from the ion continuity
we obtain
 \be
\frac{n_{i1}}{n_{i0}} =\left(\frac{\omega_{*e}}{\omega}  +
\frac{k_z^2 c_s^2 }{\omega^2} - k_y^2 \rho_s^2 \right) \, \frac{e
\phi_1}{\kappa T_e}, \quad \rho_s=c_s/\Omega_i, \quad c_s^2= \kappa
T_e/m_i. \label{e8} \ee The dispersion equation within the
quasi-neutrality limit  reads
\be \frac{\omega_{*e}}{\omega}  + \frac{k_z^2 c_s^2 }{\omega^2} -
k_y^2 \rho_s^2 =   \frac{\omega_{*e} + i D_p + i D_z (\omega^2 +
\nu_{ne}^2)/(\omega^2- i \nu_{ne} \omega)}{ \omega  + i D_p + i D_z
(\omega^2 + \nu_{ne}^2)/(\omega^2- i \nu_{ne} \omega)}. \label{e9}
\ee The term with the sound speed is due to the ion parallel
response  and it is  known to introduce the threshold for the
instability.$^{6}$

For negligible collisions, the right-hand side in Eq.~(\ref{e9})
reduces to 1, and we have the simple expression for the drift wave
in an ideal plasma \be
\omega^2(1+ k_y^2 \rho_s^2) - \omega_{*e} \omega - k_z^2 c_s^2=0.
\label{id}
 \ee
It has two solutions, one positive coupled drift-acoustic mode, and
one  negative acoustic mode.

\begin{figure}
\includegraphics[height=6.5cm, bb=0 0 520 440, clip=]{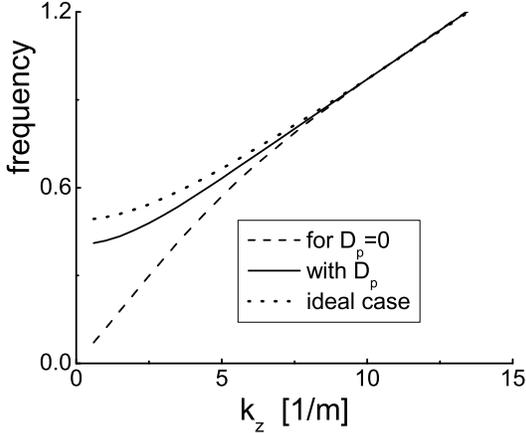}
\vspace*{-5mm} \caption{The positive solutions of Eq.~(\ref{e9})
without (dashed line) and with (full line)  the perpendicular
electron collisions included. Dotted line is the positive solution
of Eq.~(\ref{id}). Frequency is normalized to $\omega_{*e}$.
}\label{fig1}
% \vspace{0.3cm}
\end{figure}

The full Eq.~(\ref{e9}) is solved numerically for several values of
the neutral density. In Fig.~1 the frequency (normalized to
$\omega_{*e}$) of the positive drift-acoustic solution is given in
terms of the parallel wave-number  $k_z$, for a fixed $k_y=10^3$
1/m, and for $n_{n0}=10^{21}$ m$^{-3}$, $n_{e0}=5 \cdot 10^{16}$
m$^{-3}$, $T_e=11600$ K, $B_0=0.1$ T, and $L_n=0.1$ m.  This
particular set of parameters can easily be achieved in laboratory
conditions, and it is taken in order to have perfectly satisfied all
conditions used in the model. This is seen from the following.
$\omega_{*e}\simeq 10^5$ Hz, and the mode frequency is below
$\omega_{*e}$ or of the same order, so that for the given parameters
both electrons and ions are magnetized, $\Omega_i\simeq 10^7$ Hz,
$\Omega_e= 2 \cdot 10^{10}$ Hz. The electron collision frequency
with ions $\nu_{ei}$ is about 3 orders of magnitude below
$\nu_{en}\simeq 10^8$ Hz, and it is therefore negligible. The plasma
$\beta$ is about $10^{-6}$ and the electrostatic limit is justified.
Large $k_y$ implies a reasonably well satisfied local analysis
condition because $L_n/\lambda_y\simeq 16$. In the range of
frequencies and $k_z$ given in Fig.~1, the ratio $(\omega/k_z)/\vte$
changes between 0.12 (for $k_z=1$) and 0.02 (for $k_z=14$), so that
neglecting the left-hand side of the electron momentum equation is
justified. Observe also that the electron mean free path is
$\vte/\nu_{en}\simeq 0.004$ m, while the shortest parallel
wavelength is about 0.45~m,  so the condition for the fluid model
for electrons is well satisfied.

\begin{figure}
\includegraphics[height=6.5cm, bb=0 0 560 420, clip=]{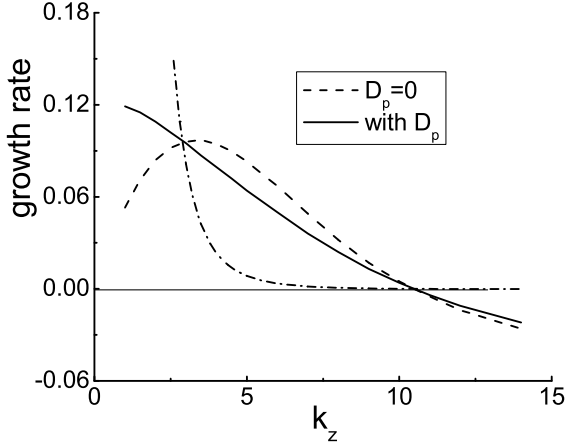}
\vspace*{-5mm} \caption{ Growth rate (normalized to $\omega_{*e}$)
associated with the mode from Fig.~1. The dash-dot line refers to
the growth rate which follows from Eq.~(\ref{c3}), i.e.,  in the
limits $D_p=0$ and $|\omega| \ll D_z$).
  }\label{fig2}
% \vspace{0.3cm}
\end{figure}

The dotted line in Fig.~1 is the solution which follows from the
ideal case (\ref{id}) (i.e., for  the right-hand side in
Eq.~(\ref{e9}) equal to unity). In the limit of small $k_z$ this
solution  goes to $0.49\,\omega_{*e}$ because here $k_y
\rho_s=1.02$. The dashed line is the numerical solution of
Eq.~(\ref{e9}) in the case when the electron collisions in the
perpendicular direction are simply ignored ($D_p=0$). The full line
is the solution including  $D_p$. Observe the significant difference
in the limit of small  $k_z$ between the ideal mode frequency  and
the collisional modes. The effect of keeping $D_p$ here is
practically to restore (at least quantitatively) the drift-like
behavior of the mode.

The growth-rates associated with the frequency from Fig.~1 are
presented in Fig.~2, showing an angle dependence and having  a
maximum at $k_z$   around $3$ (for the case $D_p=0$). We note that a
similar angle dependence (though due to ion collisions) was obtained
recently in the problem of current driven acoustic modes.$^{7,17}$
In the present case, the origin of the  maximum,   and of the
reduced frequency for small $k_z$  is different and can be
understood  from the following. If $D_p$ is completely omitted, the
approximative growth rate is given   by$^{6}$ \be
\omega_i\simeq \frac{\omega_r^2}{D_z} \frac{\omega_r^2 k_y^2
\rho_s^2 - k_z^2 c_s^2}{\omega_r^2 (1+ k_y^2 \rho_s^2) + k_z^2
c_s^2}. \label{c3}
 \ee
Here $\omega_{r, i}$ denote the real and imaginary parts of the
frequency, respectively. This  analytic expression is obtained and
valid within the approximation \be
|\omega_r| \ll D_z. \label{c1} \ee Hence, in this limit,  the growth
rate appears to be proportional to $1/D_z\sim 1/k_z^2$ and it
strongly grows for small $k_z$. This is presented by the dash-dotted
line in Fig.~2.

The condition (\ref{c1}) can be written as
 \be
\frac{|\omega_r|/k_z}{\vte} \ll \frac{k_z \vte}{\nu_{en}}.
\label{c2} \ee The right-hand side is the ratio of the mean free
path and the parallel  wavelength, and it must be  below 1 to have a
proper fluid theory. The left-hand side (the ratio of the parallel
wave phase speed and the electron thermal speed) must  also  be
below 1 within the model used here (neglected left-hand side of the
electron momentum equation).  However, although both sides
separately have some physical meaning and both  are below 1, their
mutual ratio  is arbitrary  and, therefore, the conditions
(\ref{c1},\ref{c2}) do not have to be satisfied. In fact, in our
present case, for the parameters used above and in the limit of
small $k_z$, the frequency  $|\omega_r|$ becomes of the order of $
D_z$ or even larger. This reverses  the direction of the dash-dotted
line in Fig.~2  for small enough $k_z$, and we obtain the dashed
line (the case $D_p=0$). Hence, for  small $k_z$ the growth rate
takes a completely different shape if the condition (\ref{e1}) is
satisfied, and if the dispersion equation is solved for parameters
when the condition (\ref{c1}) does not hold.

Remark that the condition (\ref{c1}) is usually used for analytical
convenience. We have shown here that this excludes an important
limit in which the growth rate may be drastically reduced.

In addition to this,  in the limit opposite to  the condition
(\ref{e1}), the perpendicular electron dynamics and perpendicular
collisions must be taken into account. As a result the growth rate
is additionally modified and it is given by the full line in Fig.~2.
This effect appears because for $k_z<6$ 1/m, we have  $D_p>D_z$ as
seen from Fig.~3, where $D_p$ is presented normalized to local
values of $D_z(k_z)$. Basically the same reasons are behind the
reduced frequency in Fig.~1.

\begin{figure}
\includegraphics[height=6.5cm, bb=0 0 560 420, clip=]{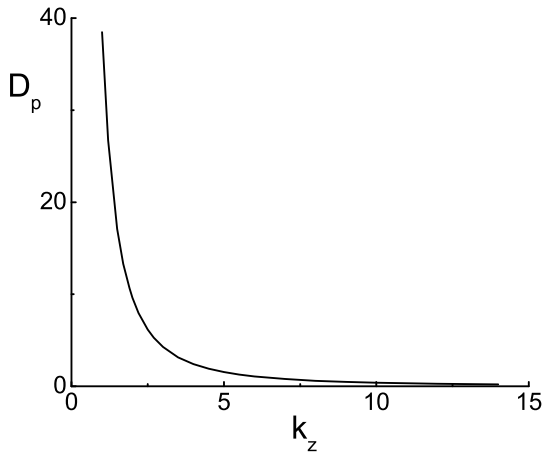}
\vspace*{-5mm} \caption{ $D_p=\nu_{en} k_y^2 \rho_e^2$ normalized to
$D_z$ in terms of $k_z$ corresponding to Figs.~1,~2.  }\label{fig3}
% \vspace{0.3cm}
\end{figure}

Note also that the growth rate in Fig.~2 becomes negative around
$k_z\simeq 10.45$ 1/m. The reason for this may be seen from
Eq.~(\ref{c3}), where the growth rate changes the sign if
$k_z^2>\omega_r^2 k_y^2 \rho_s^2/c_s^2$. This is the instability
threshold mentioned earlier, due to the parallel ion response. The
point of the sign change is almost the same for all three  cases
from Fig.~2 because  all parameters, except the frequency,  are kept
constant, and the frequency itself  for the three cases is almost
the same (see Fig.~1).

\begin{figure}
\includegraphics[height=6.5cm, bb=0 0 560 420, clip=]{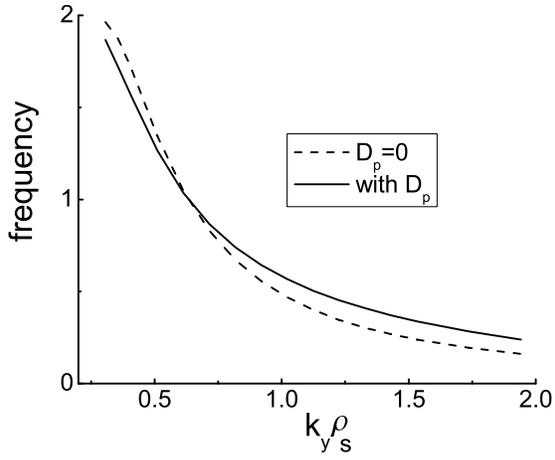}
\vspace*{-5mm} \caption{The normalized frequency as positive
solution of Eq.~(\ref{e9}) in terms of $k_y$. }\label{fig4}
% \vspace{0.3cm}
\end{figure}

\begin{figure}
\includegraphics[height=6.5cm, bb=0 0 560 420, clip=]{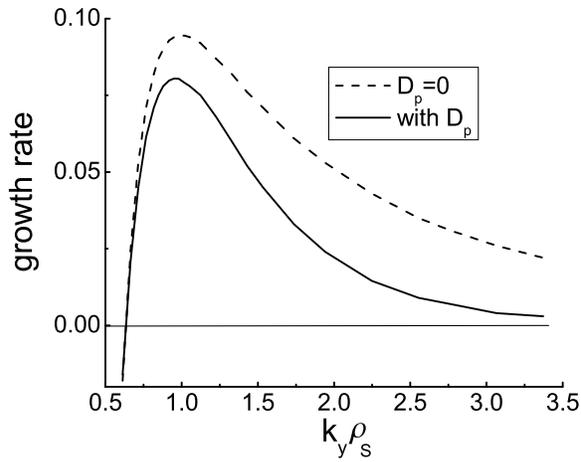}
\vspace*{-5mm} \caption{ The normalized growth rate for the
frequency from Fig.~4. }\label{fig5}
% \vspace{0.3cm}
\end{figure}

Equation~(\ref{e9}) is solved also in terms of $k_y$, for a fixed
$k_z=4$ m$^{-1}$  and for the same values for the other parameters
as above.  This yields  $D_z=25810$ Hz and $\rho_s=10^{-3}$ m. The
result for the  frequency and the growth rate (both in units of
$\omega_{*e}$) is presented in Figs.~4,~5. In the given range of
$\vec k$ the ratio $D_p/D_z$ has  values  in the range from $0.2$
(for $k_y=300$ 1/m) to $15$ (for $k_y=2500$). The growth rate
changes the sign at $k_y \rho_s \simeq 0.65$. Setting $D_p=0$
provides the dashed lines in Figs.~4 and 5. The growth rate  is
lower in the presence of $D_p$. Physically,  in the situation with
$D_p\neq 0$,  light electrons have more possibility to shield
electrostatic perturbations by moving now also in the perpendicular
direction (due to the collisions). As a result, the growth rate is
reduced.

\vspace{0.4cm}

\noindent{\bf B. Acoustic  wave}

\vspace{0.4cm}

In the case of high wave frequencies
\be \Omega_e\gg |\omega| \gg \Omega_i, \label{ia1}
 \ee
the ions are unmagnetized and their response is the same as in an
ion acoustic mode, while the electron dynamics is the same as in the
previous text. The ion continuity and momentum equation (\ref{i1})
without the Lorentz force instead of Eq.~(\ref{e8})  yield
$n_{i1}/n_{i0}=(k^2 c_s^2/\omega^2) e \phi_1/(\kappa T_e)$,
$k^2=k_y^2 + k_z^2$. The dispersion equation now reads
\be
 \frac{k^2 c_s^2 }{\omega^2}  =   \frac{\omega_{*e} + i D_p + i D_z (\omega^2 + \nu_{ne}^2)/(\omega^2- i \nu_{ne} \omega)}{
\omega  + i D_p + i D_z (\omega^2 + \nu_{ne}^2)/(\omega^2- i
\nu_{ne} \omega)}. \label{e9a} \ee Equation~(\ref{e9a}) is solved
for modified parameters in order to have the conditions (\ref{ia1})
satisfied. Hence, we take heavier (argon) ions $m_i=40 m_p$,
$m_n=m_i$, and $T_e=5$ eV, $n_{n0}=10^{21}$ m$^{-3}$, $n_{e0}=
n_{i0}=5 \cdot 10^{16}$ m$^{-3}$, $B_0=0.01$ T, $L_n=0.1$ m. For
these parameters we have$^{18}$ $\sigma_{en}=8.7\cdot 10^{-20}$
m$^2$. We set $k_y=7\cdot 10^2$ 1/m and solve Eq.~(\ref{e9a}) in
terms of $k_z$. The results are given in Figs.~6 and 7. It is seen
that the perpendicular electron collisions drastically destabilize
the mode. In the limit of small $k_z$, the growth rate in Fig.~7 is
about 70 times larger. Note that for $k_z=0.3$ 1/m we have
$D_p/D_z=141$, while for $k_z=14$ 1/m this ratio is only 0.06.

For large $k_z$ values, the growth rate in Fig.~7 decreases with
$k_z$ similar to the solutions from Ref.~2. It can also be shown
that the growth rate becomes larger when $n_{n0}$ is increased
(i.e., for larger collision frequency). These features are in
agreement with an approximative analytical solution that may be
derived from Eq.~(\ref{e9a}) assuming a dominance of the $D_z$ term.

\begin{figure}
\includegraphics[height=6.5cm, bb=0 0 560 420, clip=]{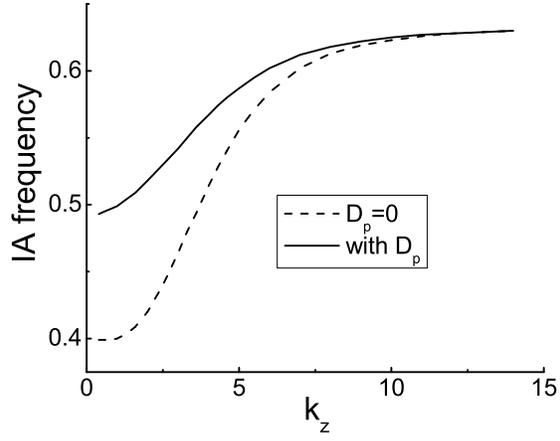}
\vspace*{-5mm} \caption{Ion acoustic frequency (normalized to
$\omega_{*e}$) as the positive solution of Eq.~(\ref{e9a}).
}\label{fig6}
% \vspace{0.3cm}
\end{figure}

\begin{figure}
\includegraphics[height=6.5cm, bb=0 0 560 420, clip=]{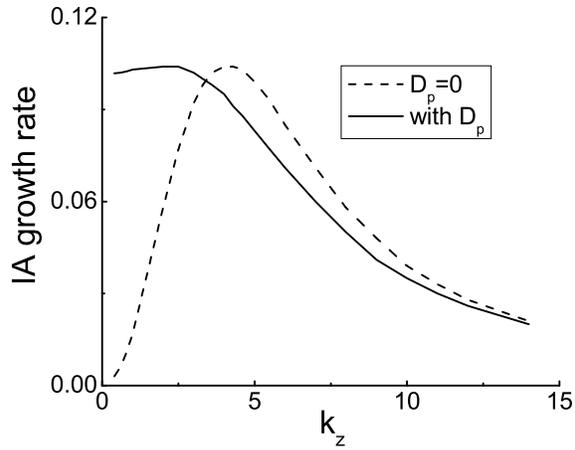}
\vspace*{-5mm} \caption{ The normalized growth rate associated with
the acoustic mode from Fig.~6.   }\label{fig7}
% \vspace{0.3cm}
\end{figure}

%\pagebreak

\vspace{0.7cm}

\noindent{\bf IV. \,\,\, CONCLUSIONS }

\vspace{0.7cm}

To summarize, we have presented some details of the collisional
drift wave instability that are usually overlooked in the
literature. In particular, the mode behavior for small parallel wave
numbers is usually described within the limit given by the condition
(\ref{c1}), resulting in a strongly increasing growth rate
proportional to $1/k_z^2$. We have shown that in the opposite limit
the mode is considerably modified, both the frequency and the growth
rate are reduced. As a result, the growth rate appears to be  angle
dependent, reaching a maximum value for a certain angle of
propagation. Furthermore,  the effect of collisions on the
perpendicular electron dynamics and on the collisional drift wave
instability is  discussed. It is shown to have effect on  both the
wave frequency and the growth rate, as presented in Figs.~1,~2,~4,
and 5.

Similar effects of the perpendicular electron collisions are shown
for the acoustic-type mode in plasmas with magnetized electrons and
unmagnetized ions. In such plasmas, the ion electrostatic
perturbations result in  acoustic modes regardless of the angle of
propagation. The physics of the mode is described in the classic
Ref.~1, and was also discussed more recently in Ref.~2.

The collision frequency has been taken constant in the present work,
following standard semi-empirical models used in the literature, yet
a recent study$^{19}$  shows that it can be significantly modified
depending on ion temperature. This is an issue that requires further
investigations. In any case we believe that the results presented
here should be taken into account  in  studies dealing with
dissipative drift-type instabilities, like those in Refs.~20-23. A
most obvious example where the instability conditions will clearly
be modified is the recent Ref.~24 dealing with drift dissipative
instability in a toroidal device.

 The model and results obtained here may be used also in
application to space plasmas. A most obvious example where the
theory works  is the lower solar atmosphere with the ionization
ratio which may be as low as $10^{-6}$  (in photosphere, with
typical temperatures of around $5\cdot 10^{3}$ K), and it changes
with the altitude to become of the order of 1 in the chromosphere
(at the altitude of about 2200 km), and grows  towards corona where
neutrals are absent. Some aspects of the drift waves in solar
atmosphere may be found in Refs.~25,~26. A similar situation with
altitude-dependent parameters suitable for the drift wave  may be
found in the terrestrial ionosphere.$^{27}$ Another interesting
aspect of the application to space problems may be seen also in
Ref.~28 dealing with Hall thrusters used in spacecrafts propulsion,
with dominant collisions with neutrals and a geometry suitable for
the drift wave analysis.

We have briefly discussed the issue of the equilibrium in such a
three-component collisional plasma. One important point that follows
from it, and that   has  been discussed  briefly, is the possibility
of equilibrium macroscopic flows of neutrals in a magnetized plasma.
Such flows  may be generated by diamagnetic drifts of plasma species
in an inhomogeneous plasma. The diamagnetic drift itself is not a
macroscopic plasma flow but the effect of gyromotion of plasma
species in the presence of a density gradient. Yet, as such it may
still generate real macroscopic flows of the neutral component of
the plasma. This is due to friction and may have important
implications, in particular for space plasmas. Similar effects
should appear in the presence of a magnetic field gradient.

\vspace{1cm}

\paragraph{Acknowledgements:}
The  results presented here  are  obtained in the framework of the
projects G.0304.07 (FWO-Vlaanderen), C~90205 (Prodex), and
GOA/2009-009 (K.U. Leuven).

\pagebreak

\end{document}